\documentclass[aps,prb,twocolumn,showpacs,floatfix,superscriptaddress]{revtex4-2}
\usepackage{amsmath,amsxtra,amssymb,latexsym,amscd,amsthm}
\usepackage{indentfirst}
\usepackage[mathscr]{eucal}
\usepackage{graphicx}
\usepackage{natbib}
\usepackage{color}
\usepackage{mathtools}
\usepackage{xcolor}
\usepackage{titlesec}
\usepackage{mdframed}
\usepackage{amsmath}
\usepackage{natbib}
\usepackage{amssymb}

\newmdenv[linecolor=black,skipabove=\topsep,skipbelow=\topsep,
leftmargin=-10pt,rightmargin=-10pt,
innerleftmargin=10pt,innerrightmargin=10pt]{mybox}

\begin{document}
\title{ Motions of a homopolar motor inside a conducting tube}

\author{Anh Q. Do}
\affiliation{Hanoi-Amsterdam High School for the Gifted, Hanoi 10000, Vietnam}
\email{dqa100606@gmail.com}

\author{Duy V. Nguyen}
\affiliation{Phenikaa Institute for Advanced Study, Phenikaa University, Hanoi 12116, Vietnam}
\email{duy.nguyenvan@phenikaa-uni.edu.vn}
\affiliation{Faculty of Computer Science, Phenikaa University, Hanoi 12116, Vietnam}

\begin{abstract}
	We analyze the physics of a type of homopolar motor comprising an AA battery with two cylindrical neodymium magnets on each end that roll inside a metal cylindrical tube. The motion of the motor results from the interaction between the magnetic field of the magnets and the magnetic field created by the current inside the magnets. We develop a model to describe the dynamics of the system, including the calculation of the terminal velocity of the motor due to eddy currents.
	
\end{abstract}

\maketitle
\section{Introduction}
In 1821, a year after Hans Christian Oersted discovered electromagnetism, Michael Faraday figured out how to turn it into motion, leading to the inception of the world's first electric motor which nowadays is referred to as a homopolar motor. The motor consists of a conducting disc, free to rotate in the neighborhood of a permanent magnet.  In 1822, André-Marie Ampère created a new kind of motor, when he succeeded in spinning a cylindrical magnet account its axis by connecting it to a battery generating a steady current. Nowadays, it is easy to present such a motor in the classroom utilizing a strong magnet, an AA batterry, and a piece of copper wire. Over the years, a large number of publications have appeared that describe the classical Faraday and Ampère homopolar motors \cite{Guala2002,Chaib2012, Chiaverina2004, Wong200902, Wong2009}, and that provide the application of eddy currents, including electromagnetic braking, and the curious behavior of a magnetic falling inside a conductive non-magnetic metal tube\cite{Irvine2014,Donoso2011,Land2016}. A recent YouTube video showcases a homopolar motor, comprising two cylindrical neodymium magnets affixed to the ends of a battery, which rolls inside a conducting tube, and the entire system moves on a horizontal table\cite{youtube2016}. In this article, we will analyze the dynamics of this system.

The magnets are nickel-coated, and the conductive inner surface of the cylindrical metal tube enables the formation of a closed circuit between the two battery terminals when the magnets make contact with the tube's surface. Once the magnets come into contact with the tube's surface, it generates an electric current that travels through both the magnets and the tube's surface. Torques act on each of the cylindrical magnets due to the interaction of the current with the magnetic field of the cylindrical magnets, causing the rotation of the magnets. The rotating motor causes the cylinder to moves on a horizontal table surface. Once the cylindrical magnets start to rotate, it will generate a back electromotive force in the motor due to the changing magnetic flux through the rotating cylindrical magnets. Therefore, the magnitude of the current in the closed circuit decreases over time and approaches a fixed value. This reduction in current results in a gradual decrease in the motor's rotation speed until it stabilizes in a steady rotational state.

The Lorenz force and the torque acting on the cylindrical mangnets have been presented in a number of papers\cite{Stewart2006,Criado2016},  and more detail discussion in the book of Andrea Macchi, Giovanni Moruzzi, and Francesco Pegoraro\cite{Andrea2017}, they showed that the torque depends only on the current and magnetic field inside the cylindrical magnets. We derived the equations of motion for the magnets and the metal cylinder by utilizing the previously calculated results of magnetic force and torque.

\section{Theory}

\subsection{Force on the motor due to eddy current}

\begin{figure}[h!]
	\centering
	\includegraphics[scale=0.5]{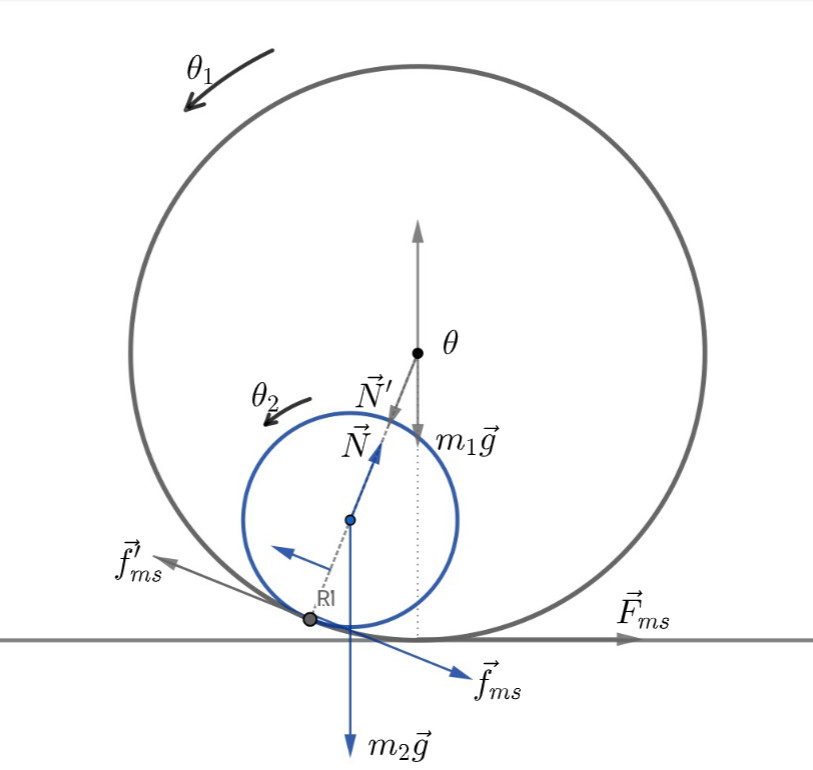}
	\caption{Schematic diagram of the homopolar motor inside a conducting tube}
	\label{fig:fig_1}
\end{figure}

Consider two indentical short cylindrical and strong magnets of radius $r_2$ attached to the  ends of the battery, roll inside a thin metal tube of radius $r_1$ (see FIG. \ref{fig:fig_1}). For a cylindrical magnet of radius of $r_2$, length $l$, and uniform magnetization $\vec M$, the uniform magnetic field inside is given by\cite{cheng1989}
\begin{equation}
	{\vec B_0} = \displaystyle \frac{{{\mu _0}\vec M}}{2}\frac{l}{{\sqrt {{{(l/2)}^2} + r_2^2} }}.
\end{equation}
This magnetic field is perpendicular to the two sides of the cylindrical magnet. As the magnets rotate, the changing magnetic fulx $\Phi_B(t)$ generate an inducted emf $\mathcal{E}$ inside the magnets. This emf can be obtained using Faraday's law,

\begin{equation}
	\mathcal{E}=- \frac{{d{\Phi _B}}}{{dt}} =  - \frac{d}{{dt}}\int {\vec B \cdot d\vec S} ,
\end{equation}
which for a motional induced emf gives the relation\cite{Donoso_2009},
\begin{equation}
	\mathcal{E}= \int {(\Vec{v}\times\Vec{B}) \cdot d\vec l}  =\int_0^{r_2}(\Vec{v}\times\Vec{B}).d\Vec{r} = \frac{B_0 \omega r_2^2}{2},
\end{equation}
where $\omega$ is the angular speed of the magnets. The current $I$ through the magnets and the outer cylinder is therefore given by
\begin{equation}
	\label{eq:eq4}
	I = \frac{{V - 2{\cal E}}}{R_T} = \frac{V}{R_T} - \frac{{{B_0}\omega r_2^2}}{R_T},
\end{equation}
where $V$ is the open-circuit voltage of the battery, and $R_T$ is the effective total resistance (which includes the internal resistance of the battery and the resistance between the contact points of the magnets with the outer cylinder).  The Lorentz force acting on the magnet is  given by, 
\begin{equation}
\vec F = I\int {d\vec l \times \vec B} ,
\end{equation}
while the torque acting on the magnet is, 
\begin{equation}
	\vec \tau  = I\int {\vec r \times (d\vec l \times \vec B)} .
\end{equation}
Assume that the current path within the magnet follows its surface. Additionally, we consider all current to follow straight-line paths. Therefore, the current flows from the lower rim to the magnet's edge and then moves radially inward toward the center of the magnet. The magnitude of the total magnetic  force acting on the motor is\cite{Andrea2017}, therefore, 
\begin{equation}
	\label{eq:eq7}
	F = 2\int_0^{r_2} B I dr = \frac{2 V B_0 r_2}{R_T} - \frac{2 B_0^2 \omega r_2^2}{R_T},
\end{equation}
where we have used the current $I$ given by Eq. (\ref{eq:eq4}). The magnitude of the total torque acting on the motor is\cite{Andrea2017}:
\begin{equation}
	\label{eq:eq8}
	\tau = 2\int_0^{r_2} B I r dr = \frac{V B_0 r_2^2}{R_T} - \frac{B_0^2 \omega r_2^3}{R_T}.
\end{equation}
Equations (\ref{eq:eq7}) and  (\ref{eq:eq8}) show  that the motor rotates faster, the magnetic force and torque decrease. Therefore, the speed of the motor increases over time and approaches a fixed value. This particular value  is calculated  in the next section.
%
%
%
%


\subsection{Dynamics of the homopolar motor}
Consider a homopolar motor of mass $m_2$, moment of inertia $\mathcal{I}_2$, rolls without slipping inside a metal cylinder of mass $m_1$, and moment of inertia $\mathcal{I}_1$.  We utilize a left-handed reference frame with the origin at $O$, where the $x$-axis lies horizontally, and the $y$-axis extends opposite the direction of gravity. Let $\theta_2$, $\theta_1$,   and $\theta$ be the angular postions of the magnets, the metal cylinder,  and the center of mass of the magnets, respectively. We assume that the center of mass of the motor lies on the symmetry axis of the motor, and the metal cylinder  rolls without slipping on the horizotal table. Therefore, 
\begin{align}
	\label{eee5}
	 \theta_1r_1  &= x,\\
	 \label{eee6}
	 \theta_2r_2&=\theta_1r_1 + \theta(r_1 - r_2),
\end{align}
where $x$ is the position of the center of mass of the metal cylinder. Let $\Vec{a}_2$ and $\Vec{a}_1$ be the acceleration of the center of mass of the motor and the metal cylinder. The relative acceleration $\Vec{a}_{2/1} =\Vec{a}_2 - \Vec{a_1}$ between the magnets and the metal cylinder is 
\begin{equation}
	 {a}_{2/1}=\ddot \theta (r_1 - r_2).
\end{equation}
The Newton's second law for the translational center of mass motion of the motor and the outer cylinder is:
\begin{align}
	\label{eq:eq12}
	m_2\Vec{a}_2 &= m_2\Vec{g} + \Vec{N} + \Vec{f}_{\rm{ms}} + \Vec{F}, \\
	\label{eq:eq13}
	m_1\Vec{a}_1 &= m_1\Vec{g} + \Vec{N}'+ \Vec{N}_1 + \Vec{f}'_{\rm{ms}} + \Vec{F}_{\rm{ms}},
\end{align}
here,  $\Vec{f}_{\rm{ms}}$ and $\Vec{f}'_{\rm{ms}} = - \Vec{f}_{\rm{ms}}$ are the frictional forces exerting on the surface of the magnets and the outer cylinder, respectively, while $\Vec{F}_{\rm{ms}}$ is the frictional force exerted by the horizontal table on the outer cylinder and  $g$ is the gravitational acceleration. The normal forces $\Vec{N}$, and $\Vec{N}' = -\Vec{N}$ at the point of contact between the magnets and the outer cylinder lie in the direction toward the center of mass, while $\vec{N}_1$ is the normal force on the metal cylinder from the horizontal table. The total magnetic forces $\Vec{F}$ is given by Eq. (\ref{eq:eq7}). Assume that the effects of air resistance on the magnets and the outer cylinder are negligible. For rotational motion about the axis of symmetry of the motor and the outer cylinder, the torque equations are
\begin{align}
	\label{eq:eq14}
	\mathcal{I}_2\ddot \theta_2 &= f_{\rm{ms}}r_2 - \tau, \\
	\mathcal{I}_1\ddot \theta_1 &= F_{\rm{ms}}r_1 -  f_{\rm{ms}}r_1 ,
	\label{eq:eq15}
\end{align}
where the total torque $ \tau $ acting on the motor is given by Eq. (\ref{eq:eq8}). The vector equation (\ref{eq:eq12}) can now  be decomposed into the following two component equations:
\begin{align}
	\label{eq:eq16} 
	m_2\left[\ddot x\cos\theta + \ddot\theta(r_1 - r_2) \right] &= F - m_2g\sin\theta - f_{\rm{ms}}  , \\
	\label{eq:eq17}
	m_2\ddot x\sin\theta &= m_2g\cos\theta - N 	,
\end{align}
where we have used $\vec{a}_2 = \vec{a}_1 + \vec{a}_{12}$, ${a}_{2/1}=\ddot \theta (r_1 - r_2)$ and $a_1 = \ddot  x$. The vector equation (\ref{eq:eq13}) is decomposed into one component in the $x$-direction:
\begin{equation}
	\label{eq:eq18}
	m_1 \ddot x = N\sin\theta + f_{\rm{ms}}\cos\theta - F_{\rm{ms}}.
\end{equation}
Using Eqs. (\ref{eq:eq14}) and (\ref{eq:eq15}) gives us:
\begin{align}
	\label{eq:eq19}
	f_{\rm{ms}} &= \frac{\tau}{r_2} + \frac{\mathcal{I}_2}{r_2^2} \left[\dot v + \ddot\theta(r_1 - r_2)\right],  \\
	\label{eq:eq20}
	F_{\rm{ms}}  &= \frac{\mathcal{I}_1}{r_1^2}\dot v + \frac{\tau}{r_2} + \frac{\mathcal{I}_2}{r_2^2} \left[\dot v + \ddot\theta(r_1 - r_2)\right], 
\end{align}
where we used  $ \ddot \theta_2 = [\ddot \theta_1r_1 + \ddot \theta(r_1 - r_2)]/r_2$ and $ \ddot \theta_1 = \ddot x/r_1 = \dot v/r_1$. Based on experimental observations, we assume that the angular position $\theta$ of center of mass of the magnets is small ($\theta \ll 1 $). Using approaximations
$\sin\theta \approx \theta$, $\cos\theta \approx 1$ and substituting $f_{\rm{ms}}$, $F_{\rm{ms}}$ from Eqs. (\ref{eq:eq19}) and (\ref{eq:eq20}),  $\tau$ from Eq. (\ref{eq:eq8}) into Eqs. (\ref{eq:eq17}) and (\ref{eq:eq18}), and simplifying, gives
\begin{align}
	\label{eq:eq21}
	&\left(m_1 + \displaystyle\frac{\mathcal{I}_1}{r_1^2}\right) \dot v=m_2g\theta,\\
	\label{eq:eq22}
	&\left(m_2 + \displaystyle\frac{\mathcal{I}_2}{r_2^2}\right)(r_1 - r_2)\ddot\theta = \displaystyle\frac{V B_0 r_2}{R_T} - \displaystyle\frac{B_0^2r_2^2}{R_T}\left[\dot v + (r_1 - r_2)\dot\theta\right].
\end{align}
From Eq. (\ref{eq:eq21}), the relation of the angular position $\theta$ of the center of mass of the magnets and the acceleration of the metal cylinder is
\begin{equation}
\label{eee13}
     \theta = \frac{m_1 r_1^2+ \mathcal{I}_1}{m_2g r_1^2}\dot v.
\end{equation}
Substituting this angular position $\theta$ into the Eq. (\ref{eq:eq22}) yields the following third-order differential equations for $v$ and $\theta$ :
\begin{align}
\label{eq:eq24}
    &\dddot v+   A \ddot v + C \dot v + D \left(v - \frac{V}{Br_2}\right) = 0,\\
\label{eq:eq25}
    &\dddot\theta + A\ddot\theta + C\dot\theta + D\theta = 0,
\end{align}
where we have defined
\begin{align}
	\label{eq:eq26}
    A &= \displaystyle \frac{B_0^2r_2^2}{R_T\left(m_2 + \displaystyle \frac{\mathcal{I}_2}{r_2^2}\right)}, \\
    C &= \displaystyle \frac{\left(m_1 + \displaystyle \frac{\mathcal{I}_1}{r_1^2} + m_2 + \displaystyle\frac{\mathcal{I}_2}{r_2^2}\right)m_2g}{\left(m_1 + \displaystyle \frac{\mathcal{I}_1}{r_1^2}\right)\left(m_2 + \displaystyle\frac{\mathcal{I}_2}{r_2^2}\right)(r_1 - r_2)},\\
    \label{eq:eq28}
    D &= \displaystyle\frac{B_0^2r_2^2m_2g}{R_T\left(m_2 + \displaystyle\frac{\mathcal{I}_2}{r_2^2}\right)\left(m_1 + \displaystyle\frac{\mathcal{I}_1}{r_1^2}\right)\left(m_2 + \displaystyle\frac{\mathcal{I}_2}{r_2^2}\right)}.
\end{align}
The third-order differential equations (\ref{eq:eq24}) and (\ref{eq:eq25}) have the same form
\begin{equation}
	\label{eq:eq29}
    \dddot y + A\ddot y + C\dot y + Dy = 0. 
\end{equation}
We can guess the a solution of of the form  $y(t) = y_0 e^{\lambda t}$, in which $\lambda$ is the time constant and $y_0$ is determined by the innitial conditions. Plugging   $y(t) = y_0 e^{\lambda t}$ into Eq. (\ref{eq:eq29}) and canceling the nonzero factor of $y_0 e^{\lambda t}$, gives us the following cubic equation for $\lambda$:
\begin{equation}
	\label{eq:eq30}
     \lambda^3 + A\lambda^2 + C\lambda + D = 0.
\end{equation}
Using Cardano's method to solve the cubic equation (\ref{eq:eq30}), gives us the solutions for $\lambda$\cite{nickalls_1993}:
\begin{align}
    \lambda_1 &= \frac{A}{3} - (k - n),\\
    \lambda_2 &= \frac{A}{3} + \frac{k - n}{2} + i\frac{\sqrt{3}}{2}(k + n),\\
    \lambda_3 &= \frac{A}{3} + \frac{k - n}{2} - i\frac{\sqrt{3}}{2}(k + n),
\end{align}
where we have defined $k$ and $n$ as follows
\begin{align}
    k &= \sqrt[3]{\frac{q}{2} + \sqrt{\displaystyle\frac{q^2}{4} + \displaystyle\frac{p^3}{27}}},\\
    n &= \sqrt[3]{- \frac{q}{2} + \sqrt{\frac{q^2}{4} + \displaystyle\frac{p^3}{27}}},
\end{align}
here $p$ and $q$ are expressed in terms of constants which are defined in Eqs. (\ref{eq:eq26}) - (\ref{eq:eq28}),
\begin{align}
	p &= C - \frac{A^2}{3},\\
	q &= \frac{2}{27}A^3 - \frac{AC}{3} + D.
\end{align}
In the case of  $\lambda = \lambda_2$ or $\lambda = \lambda_3$, which have the positive real parts, the velocity of the system goes to infinity for the large $t$. Therefore, we have really found only one solution $y(t) = y_0 e^{-|\lambda| t}$, where $\lambda = \lambda_1$. Using intitial condition $v=0$, we find the solution for $v$ and $\theta$ as follows
\begin{align}
	\label{eq:eq38}
    v(t) &= \displaystyle\frac{V}{B_0r_2}\left(1 - e^{- |\lambda| t} \right),\\
    \label{eq:eq39}
    \theta(t) &= \displaystyle\frac{V}{B_0r_2}\displaystyle\frac{{m_1}r_1^2 + {\mathcal{I}_1}}{m_2gr_1^2}|\lambda| e^{- |\lambda| t}.
\end{align}
Equation (\ref{eee6}) gives the angular speed of the magnets as
\begin{align}
	\label{eq:eq40}
	\omega (t) &= {\dot \theta _2} = v + \frac{{{r_1} - {r_2}}}{{{r_2}}}\dot \theta  \nonumber\\
	 &= \frac{V}{{Br_2^2}}\left[ {1 - \frac{{{m_2}gr_1^2 + \left( {{m_1}r_1^2 + {\mathcal{I}_1}} \right)({r_1} - {r_2}){\lambda ^2}}}{{{m_2}gr_1^2}}{e^{ - |\lambda |t}}} \right].
\end{align}
Equations (\ref{eq:eq39}) and (\ref{eq:eq40}) shows that the motion becomes uniform  for the large $t$. Therefore,  in the steady state, the metal cylinder reach the terminal speed
\begin{align}
	v_{\tau} &= \displaystyle\frac{V}{B_0r_2},
\end{align}
and the motor remain in the vertical position with $\theta = 0$ and rotate around its symmetry axis at a constant speed
\begin{equation}
	\omega_{\tau} = \frac{V}{{B_0r_2^2}}.
\end{equation}
Substituting the angular speed $\omega$ of the magnets from Eq. (\ref{eq:eq39}) into Eq. (\ref{eq:eq4}), gives the time evolution of the current through the circuit 
\begin{equation}
	I(t)=\frac{{{m_2}gr_1^2 + \left( {{m_1}r_1^2 + {{\cal I}_1}} \right)({r_1} - {r_2}){\lambda ^2}}}{{{m_2}gr_1^2}}\frac{V{e^{ - |\lambda |t}}}{{{R_T}}}.
\end{equation}
This current goes to zero for large values of $t$. This is also consistent with the law of conservation of energy because when the motor moves uniformly and neglects the air resistance torque, so the battery does not need to do work. The current will reach a fixed nonzero value if we take into account the effects of air resistance on the outer cylinder.

\section{Summary}
We studied the dynamics of a homopolar motor consisting of an AA battery with two cylindrical neodymium magnets on each end, which moves inside a metal cylinder rolling without slipping on a horizontal table. Using the previously calculated results of magnetic force and torque, we derived the equations of motion for the magnets and the metal cylinder. By applying Cardano's method, we found solutions for the time evolution of the cylinder's velocity and the motor's angular speed. The motion of the motor in this case shares similarities with the motion of a magnet falling through a copper tube; the motor accelerates and reaches a terminal velocity. The value calculated for this terminal velocity of the motor depends solely on the battery's voltage, the magnetic field inside the magnets, and the radius of the magnets.
 
\bibliographystyle{apsrev4-2}
\bibliography{ref}

\end{document}